\documentstyle[12pt,ncs]{article}
\def\lsim{\raise0.3ex\hbox{$\;<$\kern-0.75em\raise-1.1ex\hbox{$\sim\;$}}}
\def\gsim{\raise0.3ex\hbox{$\;>$\kern-0.75em\raise-1.1ex\hbox{$\sim\;$}}}
\def\21{$SU(2) \otimes U(1)$}
\def\321{$SU(3)  \otimes SU(2)  \otimes U(1)$}%

\let\vev\VEV

\def\eq#1{{Eq. (\ref{#1})}}

\def\np#1#2#3{           {\it Nucl. Phys. }{\bf #1} (19#2) #3}
\def\pl#1#2#3{           {\it Phys. Lett. }{\bf #1} (19#2) #3}
\def\pr#1#2#3{           {\it Phys. Rev. }{\bf #1} (19#2) #3}
\def\prep#1#2#3{         {\it Phys. Rep. }{\bf #1} (19#2) #3}

\def\n.c.#1#2#3{         {\it Nuovo Cim. }{\bf #1} (19#2) #3}
\def\r.n.c.#1#2#3{       {\it Riv. del Nuovo Cim. }{\bf #1} (19#2) #3}

\relax

\begin{document}

\baselineskip=7mm

\relax
\newcommand{\eV}{\,{\rm eV}}
\newcommand{\Tr}{{\rm Tr}\!}
\renewcommand{\arraystretch}{1.2}
\newcommand{\beq}{\begin{equation}}
\newcommand{\eeq}{\end{equation}}
\newcommand{\bea}{\begin{eqnarray}}
\newcommand{\eea}{\end{eqnarray}}
\newcommand{\ba}{\begin{array}}
\newcommand{\ea}{\end{array}}
\newcommand{\bmat}{\left(\ba}
\newcommand{\emat}{\ea\right)}
\newcommand{\refs}[1]{(\ref{#1})}
\newcommand{\ler}{\stackrel{\scriptstyle <}{\scriptstyle\sim}}
\newcommand{\ger}{\stackrel{\scriptstyle >}{\scriptstyle\sim}}
\newcommand{\lag}{\langle}
\newcommand{\rag}{\rangle}
\newcommand{\ns}{\normalsize}
\newcommand{\cm}{{\cal M}}
\newcommand{\gr}{m_{3/2}}
\newcommand{\p}{\partial}
\def\rp{ $R_P$} 
\def\321{$SU(3)\times SU(2)\times U(1)$}
\def\tl{{\tilde{l}}}
\def\tL{{\tilde{L}}}
\def\bd{{\overline{d}}}
\def\tL{{\tilde{L}}}
\def\a{\alpha}
\def\b{\beta}
\def\g{\gamma}
\def\c{\chi}
\def\d{\delta}
\def\D{\Delta}
\def\db{{\overline{\delta}}}
\def\Db{{\overline{\Delta}}}
\def\e{\epsilon}
\def\l{\lambda}
\def\n{\nu}
\def\m{\mu}
\def\nt{{\tilde{\nu}}}
\def\p{\phi}
\def\P{\Phi}
\def\x{\xi}
\def\r{\rho}
\def\s{\sigma}
\def\t{\tau}
\def\th{\theta}
\def\ne{\nu_e}
\def\nm{\nu_{\mu}}
\def\rp{$R_P$}
\def\mp{$M_P$}     
\renewcommand{\Huge}{\Large}
\renewcommand{\LARGE}{\Large}
\renewcommand{\Large}{\large}
\begin{titlepage}
\title{{\bf Gravitational  Violation  of R  Parity\\}
       {\bf and  its  Cosmological  Signatures  \\}
                         \vspace{-3cm}
   \hfill{\ns hep-ph/9608307}\\
                             \vspace{2.5cm} }

\author{ V. Berezinsky$^\dagger$, \hspace{.4cm}
         Anjan S.~Joshipura$^*$, \hspace{.4cm} 
         Jos\'e W. F. Valle$^{**}$ \\[.5cm]
  {\ns\it $^\dagger$Inst. Nazionale di Fisica Nucleare,}\\
  {\ns\it Lab. Nazionale del Gran Sasso, Assergi (AQ), Italy} \\[.3cm]
  {\ns\it $^*$Theoretical Physics Group, Physical Research Laboratory}\\
  {\ns\it Navarangpura, Ahmedabad, 380 009, India} \\[.3cm]
{\ns\it $^{**}$ Instituto de F\'{\i}sica Corpuscular - C.S.I.C.}\\
{\ns\it Departament de F\'{\i}sica Te\`orica, Universitat de Val\`encia}\\
{\ns\it 46100 Burjassot, Val\`encia, Spain}\\
{\ns\it URL: http://neutrinos.uv.es}\\
  }
\date{}
\maketitle
\vspace{0cm}
\begin{abstract} 
\baselineskip=6mm

The discrete R-parity ($R_P$) usually imposed on the 
Supersymmetric (SUSY) models 
is expected to be broken at least gravitationally. 
If the neutralino is a dark matter particle its decay channels into positrons,
antiprotons and neutrinos are severely constrained from astrophysical 
observations. These constraints are shown to be violated even for
Planck-mass-suppressed dimension-five interactions 
arising from gravitational effects.
We perform a general analysis of gravitationally 
induced $R_P$ violation and identify two plausible and astrophysically 
consistent scenarios for achieving the required suppression.
\end{abstract}
\end{titlepage}
A discrete symmetry, called R-parity ($R_P$) is imposed on
the Minimal Supersymmetric Standard Model (MSSM) \cite{foot} 
to ensure proton stability. This assumption makes the Lightest 
Supersymmetric Particle (LSP) stable. The most natural candidate
for the LSP in SUSY models is a neutralino. Indeed, calculations in 
SUSY models with soft breaking terms and radiatively induced 
electroweak breaking lead to neutralino as LSP for a wide range of allowed 
parameters.  Moreover, the relic neutralino density satisfies all 
requirements for being the cold dark matter over large parameter regions. 
The hypothesis of neutralino as Dark Matter (DM) particle is 
amenable to experimental verification \cite{review2}. Such
neutralinos can be detected both directly through their elastic 
scattering off-nuclei \cite{dir} and indirectly through the products 
of neutralino annihilation \cite{sun},\cite{pos},\cite{GC}.

Although R-parity may remain unbroken if it is a remnant of a gauge symmetry
\cite{KrWi89,disy,IR91,ib2,B-L}, there is no deep theoretical reason requiring 
it to be a symmetry of nature. In fact, many models of $R_P$ violation have 
been proposed \cite{beyond}.
These typically lead to large $R_P$ violation and thus do not 
allow the neutralino to be a dark matter particle. In contrast,
if $R_P$ violation is very mild then the lightest neutralino
is unstable, but very long-lived. Naively, one would think
that a neutralino with a lifetime of the order of the age of
the universe could be a viable dark matter candidate. However,
most  neutralino decays into {\sl visible} channels, e.g. 
containing positrons, antiprotons, gamma's and neutrinos
are severely constrained 
from observations. They typically require \cite{bvenya} neutralino 
lifetimes much larger than the age of the universe. This can only be 
realized if the violation of $R_P$ is extremely weak \cite{bvenya,vvenya}.
Such a scenario was already considered in the context of a specific mechanism 
\cite{masvalle} for  spontaneous $R_P$ violation driven by a 
vacuum expectation value (VEV) for the right-handed sneutrino.
This mechanism requires very small Yukawa coupling, i.e.
fine-tuning.
A more natural way to obtain very small $R_P$ violation is
to ascribe it to gravitational interactions. 
This violation will be described here by non-renormalizable terms 
suppressed by inverse powers of the Planck mass and should thus be
naturally very small.

A somewhat surprising conclusion that emerges out of our study 
is that dimension-five interactions are in conflict with the 
astrophysical constraints by few orders of
magnitude if the neutralino provides the cold DM.

Let us first quantitatively discuss constraints on the strength
 of $R_P$  violation. The lightest neutralino $\chi$ is given by 
a superposition of wino $\tilde{W}$, bino $\tilde{B}$ and two 
Higgsinos, $\tilde{H_1}$ and $\tilde{H_2}$ as:
\beq \label{eq:chi}
\chi=Z_{\chi\tilde{W}_3}\tilde{W}_3+Z_{\chi\tilde{B}}\tilde{B}+
Z_{\chi\tilde{H}_1}\tilde{H}_1+Z_{\chi\tilde{H}_2}\tilde{H}_2.
\eeq

We can parameterize the effective $R_P$ violating interactions 
responsible for the neutralino decay in terms of the MSSM fields as 
follows:
\bea
\label{eq:d=4}
W_{eff}&=&   \lambda_1 (U^cD^cD^c)_F+\lambda_2(L L E^c)_F+ \lambda_3
(QD^c L)_F\\ \nonumber 
&+&\epsilon (LH_2)_F
\eea
The notation for the fields is standard. We suppressed  the generation 
indices. \eq{eq:d=4} includes only {\sl renormalizable} terms relevant for 
neutralino decay. However, as it will be understood below, this 
expression has wider generality.

The above interactions result in neutralino decay to three fermions.
The width for this decay depends upon whether it proceeds
through the Higgsino or gaugino component. In the former case,
\beq
\tau_{\chi}^{-1}=\lambda_i^2 Z_{\chi \tilde{H}}^2
\frac{G_F m_f^2}{192(2\pi)^3}\frac{m_{\chi}^5}{\tilde{m}_f^4}
\label{eq:Hdec}
\eeq
where $m_{\chi}$ , $\tilde{m}_f$ and $m_f$ are  neutralino, sfermion
and fermion masses, respectively. In the case of quarks the width should 
be multiplied by the number of colours.
When the decay proceeds through the bino component $\tilde{B}$ of the 
neutralino the decay width  is
\beq
\tau_{\chi}^{-1}= \lambda_i^2 Z_{\chi\tilde{B}}^2\frac{\alpha_{em}Y_{f_R}^2}
{192(2\pi)^2\cos^2\theta_W}\frac{m_{\chi}^5}{\tilde{m}_f^4},
\label{eq:Bdec}
\eeq
where $Y_{f_R}$ is hyper-charge of the right fermion and $\theta_W$ is the 
weak mixing angle.
Finally, the width of $\chi \to \nu+e^++e^-$ due to the last term of 
\eq{eq:d=4} is 
\beq
\tau_{\chi}^{-1}= \epsilon^2 Z_{\chi\tilde{H}}^2
(\frac{1}{4}+\sin^2\theta_W+\frac{4}{3}\sin^4\theta_W)
\frac{G_F^2m_{\chi}^3}{192\pi^3},
\label{eq:Hedec}
\eeq
The condition $\tau_{\chi} > t_0$ (where $t_0$ is the age of the 
universe) leads to constraints on $\lambda_i$ and $\epsilon$.
However, much more stringent limits follow from production of positrons 
in our Galaxy \cite{bvenya,vvenya}, from the diffuse gamma-radiation
\cite{bvenya,vvenya} and from neutrino-induced muons \cite{Kam92}.
From the analysis given in \cite{vvenya} it follows
that when the decay to positrons is unsuppressed as in the
present case, the strongest constraints on both $\lambda_i$ 
and $\epsilon$ follow from the observed flux of positrons 
in our Galaxy. The lower limit on neutralino lifetime 
from this flux is \cite{vvenya}
\beq
\tau_{\chi}(\chi \to e^++\mbox{anything}) > 7\times 10^{10}\sqrt{m_{100}} t_0h ,
\label{eq:l-t}
\eeq
where $m_{100}=m_{\chi}/100$ GeV and $h$ is dimensionless Hubble constant.

Using this limit and keeping in mind indirect production of positrons 
through decay of other particles, we find 
\bea
\lambda_i &<& 
4 \times 10^{-21} Z_{\chi\tilde{H}}^{-1}(\frac{\tilde{m}_f}{1~TeV})^2
(\frac{100~\mbox{GeV}}{m_{\chi}})^{9/8}(\frac{1~\mbox{GeV}}{m_f})^{1/2}
\\
\epsilon  &<& 
6 \times 10^{-23} Z_{\chi\tilde{H}}^{-1}m_{100}^{-7/4} \mbox{GeV}
\label{eq:cons}
\eea
The above estimates show that if the neutralino is a DMP the 
R-parity violating parameters $\lambda_i$ and $\epsilon$ are very 
strongly limited from above. 

Let us now turn to a systematic dimensional analysis of 
R-parity-violating operators in the MSSM.
We have already discussed 
the operators of dimension d=4.
Extremely small couplings 
are needed for the neutralino to be a DMP, which requires fine tuning.

The relevant Planck-mass suppressed operators of dimension d=5 in the MSSM 
are the following ones:

\beq
\label{eq:d=5}
\ba{ccc}
{\beta_1\over M_P}(H_1H_2^*E^c)_D   
& {\beta_2\over M_P}(QL^*U^c)_D   &  
{\beta_3\over M_P}(U^cD^{c*}E^c)_D   \nonumber\\
\frac{\beta_4}{M_{P}}(QQQH_1)_F  &
\frac{\beta_5}{M_{P}}(LH_2H_1H_2)_F \:. 
\ea
\eeq
The first three terms in the above equation contain  the F-terms
of the anti-chiral fields $ H_2^*,L^*$ and $D^{c*}$ respectively.
These are determined in the supersymmetric limit by the standard
MSSM $R_P$ conserving superpotential. Using this one
can show that their contribution to the
neutralino decay is suppressed either due to small Yukawa couplings
or kinematically, when the charged Higgs is heavier than the 
neutralino. The contribution of the fourth term with three $Q$
fields to neutralino decay can also be shown to be small.

The first four operators thus satisfy astrophysical constraints 
without need for extra suppression in the corresponding coefficient.
In contrast, the decay induced by the last term cannot be suppressed
kinematically. 
It leads to the effective  $R_P$  breaking parameter
$$ \epsilon\sim \beta_5 M_{EW}^2/M_{P}\sim \beta_5\; 10^{-15} \mbox{ GeV} $$ 
This value of $\e$ is extremely small and leads to a neutralino lifetime
longer than the age of the universe. But  it
is in conflict with the astrophysical constraints (\ref{eq:cons}) 
unless $\beta_5 \lsim 10^{-5}-10^{-7}$. 
This situation is similar to the gravitationally induced axion 
mass \cite{KLL95}, \cite{axion} where the  quantum gravitation 
corrections are not small enough to suppress it adequately.

If $1/M_{P}$ terms are forbidden (for example by some unbroken symmetry),
then $1/M_{P}^2$ terms (d=6 operators) become important. An example of such 
operator is 
\beq
\frac{\beta}{M_{P}^2}\left((LH_2)(H_1H_2)^*\right)_D
\label{eq:d=6}
\eeq
This term gives an $\epsilon$ which is about 10 orders of 
magnitude less than needed to produce observable effects.

Therefore, while in the MSSM $1/M_{Pl}^2$ terms are too small, the
  $1/M_{Pl}$ terms are 
too large and need additional suppression, i.e. small $\beta$.

If wormhole effects are responsible for the terms we are discussing, they 
can contain a topological suppression leading to very small $\beta$.  
Generically, this suppression is described by an $e^{-S}$ factor, where 
$S$ is an action of a wormhole which absorbs the $R_P$ charge. 
In the semi-classical approach $S \sim 10$. In the case of
the Peccei-Quinn symmetry such estimates give $S \sim \ln(M_{P}/f_{PQ})
\approx 16$ resulting in a suppression factor $\beta \sim 10^{-7}$, which is
needed in our case. A detailed discussion of such suppression 
factors in wormhole effects is given in \cite{KLL95}. It is shown that 
the action $S$ is connected with the size of the wormhole throat  
$R(0)$ and can vary from $S \approx 6.7$ for the naive estimate 
$R(0) \approx M_{P}^{-1}$, up to 
$8\pi^2/g_{str}^2 \approx 190$ in string inspired models.  
Thus if the action is close to its semi-classical value
the
topological wormhole effects can provide the suppression needed 
for the long-lived neutralino $\beta \sim 10^{-7} - 10^{-5}$ 
to satisfy the  astrophysical constraints.

Apart from topological wormhole effects, suppression of d=5 operators 
can occur due to some additional symmetry. Let us assume 
that there exists a singlet sector which communicates
with the MSSM sector only gravitationally through non-renormalizable 
terms in the Lagrangian. R-parity can be broken spontaneously in this 
 sector, for example, due to some $R_P$-odd field $\eta$ 
developing a non-zero VEV. $R_P$ violation can penetrate the MSSM 
sector through non-renormalizable interactions between $\eta$ and 
the MSSM fields. 
n contrast to the first case, gravity is not 
directly responsible for the breaking of $R_P$ but 
it leads to 
effective  $R_P$  violation in the observable sector through 
the presence of non-renormalizable interactions. 
We now discuss this possibility of {\sl hidden R-parity violation}
in a model independent way and then provide an example.

Let us assume the existence of an \321 singlet field $\eta$ 
coupling to the MSSM fields only through non-renormalizable terms. 
This can be achieved by a proper symmetry as we shall discuss.
There are four dimension-five operators involving $\eta$ which 
lead to $R_P$ violation: 
\beq
\ba{cc}
O_1=\frac{\alpha_1}{M_{P}}(U^c D^c D^c\eta)_F\;\;\;\;\; 
&O_2=\frac{\alpha_2}{M_{P}}(LLE^c\eta)_F\\ \nonumber
O_3=\frac{\alpha_3}{M_{P}}(QD^cH_1\eta)_F\;\;\;\;\;\;& 
O_4=\frac{\alpha_4}{M_{P}}(LH_2\eta^*)_D
\label{eq:Od=5}
\ea
\eeq 
where $\alpha_{1,2,3,4}$ are parameters of order one.
These operators conserve $R_P$ if the field $\eta$ 
is chosen odd. The non-zero  VEV $\vev{\eta}$ then breaks 
$R_P$ and leads to the effective interactions in eq.(2) with 
couplings given by ($i=$1,2,3):
\beq
\ba{cc}
\lambda_i=\alpha_i \vev{\eta}/M_P\;\;\;\;\; 
& \epsilon=\alpha_{4} \vev{F_{\eta^*}}/M_P\\
\ea
\eeq

The effective $R_P$ violation among the MSSM fields is
governed by two distinct scales. One scale is $\vev{\eta}$ and
determines the trilinear interactions of eq.(2), while the 
other scale $\vev{F_{\eta^*}}$ characterizes  SUSY breaking 
in the singlet sector and determines the bilinear term $\epsilon$. 
In general, these two scales could be quite different. The 
constraints derived in (\refs{eq:cons}) imply 
\beq
\vev{\eta} \lsim 10^{-1} \mbox{GeV}; \:
\vev{F_{\eta^*}} \lsim 10^{-2} 
\mbox{GeV}^2
\label{fcons}
\eeq

If SUSY is broken through the usual soft terms, the constraint on 
$\vev{F_{\eta}}$ can easily be satisfied, as we will show in a 
specific example. In contrast, the constraint on the trilinear 
coupling implies very small $\eta$ VEV which may be unnatural. 
To avoid this one should forbid the corresponding dimension-5 
operators, in which case the dominant $R_P$ violation would 
arise from dimension-six interactions. For example, the operator
\beq
\frac{1}{M_{P}^2}[(LH_2)(H_1H_2)\eta]_F 
\label{eq:Od=6}
\eeq
leads to
$\epsilon \sim 10^{-34} \vev{\eta}$ GeV. The effect of this term
could be observable only if $R_P$ violation in the singlet sector
occurs close to the unification scale. 

Let us now consider a concrete realization of the above scenario.
We will impose a gauged discrete symmetry in order to
prevent gravitational breaking. $R_P$ by itself can be gauged 
in a suitable extension of the MSSM but it cannot prevent the 
occurrence of a term like $LH_2\eta$ which lead to a large $R_P$
violation when $\eta$ develops a vev. We therefore
look for a more general symmetry which leads to effective $R_P$
conservation at the renormalizable level. Gauged $Z_N$ symmetry
of the type considered in \cite{IR91} provides an example. 
The $\th$ is assumed to carry the $Z_N$-charge -1.
The $Z_N$ charge of one of the observable superfields is set 
to zero by appropriate redefinition of the $Z_N$ generators. Requiring 
that the standard $R_P$ conserving couplings of the MSSM fields are 
allowed by the $Z_N$ symmetry we determine the charges of the remaining 
fields in terms of two parameters $x$ and $y$ according to:

$$\begin{array}{ccccccc}
\underline{Q} & \underline{U^c,H_1} & \underline{D^c,H_2} &
 \underline{L} & \underline{E^c}   & \underline{Y} &\underline{\eta} \\ 
0 &  x      & 2-x     & y &2-(x+y)& 2 & {N\over 2}\\ 
\end{array}$$
where we have introduced a singlet field $Y$ in addition to $\eta$,
in order to obtain $R_P$ violation.  With these charge assignments 
dimension-4 terms respect  $R_P$  and the $\eta,Y$ do not couple 
to the MSSM fields in the renormalizable Lagrangian,
as long as $x\not =2$ and $x-y \not = 0,-2, N/2$.  
The most general $Z_N$-invariant renormalizable 
superpotential in this case can be written as
\cite{f1}
\beq
\label{sup}
W=W_{MSSM} + \delta \: Y(\eta ^2 - f^2)
\end{equation} 
The above superpotential gives rise to a VEV for $\eta$ at the
supersymmetric minimum, leading to effective $R_P$ violation 
 for the MSSM fields through operators of dimensionality 
$\geq 5$. 
The choice $2+y-x=N/2$ allows the dimension-6 operator of \refs{eq:Od=6}. 
 The allowed  higher dimensional terms are given in this case by 
\bea
\label{dim5-6}
{\cal L}_{NR}=
\frac{\beta_5}{M_P}(LH_2\eta^*)_D+
\frac{1}{M_P^2} [ \delta_1 (LLE^c\eta^*)_D +\nonumber \\ 
\delta_2 (QD^cL\eta^*)_D 
+\delta_3 (LH_2\eta^*Y)_D
                     + \delta_4(LH_2 H_1H_2\eta)_F ] 
\eea
Note that the dimension-5 operator above cannot be forbidden
if the dimension-6 term in \refs{eq:Od=6} is to be allowed.
As discussed above it does not lead to large $R_P$  violation
as long as SUSY remains unbroken in the singlet sector, as
happens with the superpotential choice in \refs{sup}.
In a realistic situation soft SUSY breaking introduces 
terms which make $F_{\eta,Y}$ non-zero. 
If $\vev{\eta} \lsim M_{SUSY} \sim $ TeV then 
$\vev{F_{\eta}} \sim {2\vev{\eta}^3\delta^2 A\over m_{Y}^2}$
where $m_Y \sim A$ characterizes soft SUSY breaking.
Choosing $\delta \sim 10^{-2}, m_Y \sim 10^3$ GeV,
$A \sim 10^2$ GeV and $\vev{\eta} \lsim 100$ GeV we have
$F_{\eta} \lsim 10^{-2} \mbox{GeV}^2$ in agreement with the
constraint of \refs{fcons}.
 If $\vev{\eta}$ is much larger than $M_{SUSY}$ then
$\vev{F_{\eta}}$ would also be large and induce large $\epsilon$.
This can be prevented by a symmetry. Specifically, if the
kinetic energy terms for the singlet fields are chosen to  be 
no-scale type \cite{noscale} then 
$\vev{F_{\eta,Y}}$ vanish at the minimum of the potential and effective
  breaking of $R_P$  would arise only from the dimension-6 operator.
This operator could lead to observed signatures if $\vev{\eta}$ is
very large, near the unification scale.

The $Z_N$ symmetry introduced above can be gauged if it satisfies
discrete gauge anomaly constraints discussed in \cite{ib2}. In our 
case these are given as: 
\bea
-2 N_g+6 = k_1\;N \\ \nonumber 
N_g(y-4)+4= k_2\; N \\ \nonumber 
N_g( -7+y-x)+N/2-9= k_3\; N + \kappa \; k_4\;N/2 \nonumber
\eea
where $\kappa$ is $1 (0)$ for even (odd) $N$ and $k_{1,2,3,4}$ 
are integers. The first constraint is automatically satisfied 
for three ($N_g=3$) generations. The remaining constraints can 
also be satisfied for appropriate choices of $x,y $ and $N$.
A choice which satisfies all the anomaly constraints 
and leads to the required interactions in \eq{sup} and 
\refs{dim5-6} is given by $N=3, x=1/6$ and $y=-1/3$. 
Clearly many other choices would be possible. 

In summary, there is no deep theoretical motivation for R-parity to be 
absolutely conserved. In this paper we have analyzed the possibility that 
R-parity can be weakly violated by gravitationally induced terms 
suppressed by inverse powers of the Planck mass. 
Due to astrophysical constraints (mainly to the positron production 
in our Galaxy) this extremely weak R-parity violation is compatible 
with hypothesis that the neutralino is a DM particle only if its
lifetime is about $10^{10}$ times longer than the age of the universe. 
We analyzed gravitationally induced dimension-5 operators and 
demonstrated that they break R-parity too strongly to comply with
this constraint. 
We showed that the unstable neutralino as a DM particle is  
possible if dangerous d=5 terms are strongly suppressed.
For the MSSM case, the 
required suppression could be provided by the classical  wormhole
action $S\approx 7$. 

Another possibility for very weak R-parity breaking can be provided by 
existence of additional gauge discrete symmetry. We constructed a model
with a gauge $Z_N$ symmetry and $SU(3)\times SU(2)\times U(1)$ singlet field 
$\eta$, which communicates with the MSSM fields only through gravity. R-parity 
violation is driven by VEV of this field.

The decaying neutralino has interesting astrophysical signatures. 
In some models \cite{vvenya} the neutralino decay to the Majoron J, 
$\chi \to \nu + J$ may be dominant, resulting in a detectable 
isotropic flux of mono-energetic neutrinos.
In the more general case of $R_P$ breaking by d=5 operators discussed above, 
the neutralino decay signature is weaker and is given by the ratio of 
the signals from the Sun and Earth to that from the Galactic halo.  
The signal from annihilation of neutralinos in the Earth and the Sun is 
the same as for a stable neutralino, while the positron and antiproton 
fluxes from the Galactic halo could be strongly enhanced due to neutralino 
decay.

\vskip .3cm
This work was supported by DGICYT Grants PB92-0084
and SAB94-0014 and by a CICYT-INFN grant. We thank 
Graham Ross and Mikhail Shifman for interesting discussions.


\begin{thebibliography}{99}
\bibitem{foot}
The MSSM is defined here as a supersymmetrized \321 model.
\bibitem{f1}  
The $\eta^2$ term allowed by the symmetry can be
removed by a redefinition of the field $Y$
\end{thebibliography}
\end{document}